\documentclass[aps,pra,twocolumn]{revtex4}
\usepackage{amsmath}
\usepackage{amssymb}
\usepackage{epsfig}
\usepackage{bm}

\def\be{\begin{equation}}
\def\ee{\end{equation}}
\def\bea{\begin{eqnarray}}
\def\eea{\end{eqnarray}}

\begin{document}

\title{Comment on ``Probing gravitational wave polarizations with signals
from compact binary coalescences"}
\author{Anatoly A. Svidzinsky}
\affiliation{Department of Physics \& Astronomy, Texas A\&M University, College Station,
TX 77843}
\date{\today }

\begin{abstract}
In a recent paper \textquotedblleft Probing gravitational wave polarizations
with signals from compact binary coalescences\textquotedblright\
(arXiv:1710.03794 [gr-qc]) the authors argue that a single detection of
gravitational wave by the LIGO-Virgo network is capable to distinguish
between pure tensor and pure vector polarizations of gravitational waves.
Here we point out a mistake in the author's analysis and show that such
differentiation is possible only in the unlikely event when gravitational
wave propagates in the direction of interferometer zero response for the
tensor or vector polarizations. Nevertheless, the LIGO-Virgo network can
distinguish between pure tensor and pure vector polarizations by collecting
statistics, as we showed in Phys. Scr. 92, 125001 (2017).
\end{abstract}

\maketitle

Recently, joint detection of gravitational waves by twin LIGO
interferometers in the US and Virgo interferometer in Italy became a reality 
\cite{Abbo17,Abbo17a}. This achievement opens a perspective to measure
polarization of gravitational wave and, thus, distinguish between, e.g.,
tensor \cite{Eins15} or vector \cite{Svid17} nature of gravity.

A possibility of using networks of ground-based detectors to directly
measure polarization of gravitational waves from compact binary coalescences
has been recently investigated by Isi and Weinstein \cite{Isi17}. In
particular, the authors argued that the current Advanced-LIGO-Advanced-Virgo
network can already be used to distinguish between pure tensor and pure
vector polarizations of gravitational waves even based on a single detection
event. Moreover, using such analysis it has been concluded that the first
three-detector observation of gravitational waves from a binary black hole
coalescence (GW170814)\textit{\ }favors the purely tensor polarization of
gravitational waves against purely vector \cite{Abbo17}. Next we explain why
a single detection of gravitational wave by the LIGO-Virgo network is not
sufficient to make a decisive conclusion in favor of one of the two
polarizations and point out a mistake in Isi and Weinstein's analysis of
Ref. \cite{Isi17}.

To be specific, we consider a toy example discussed by Isi and Weinstein in
Section III.A of their paper \cite{Isi17}. Namely, we consider an
elliptically-polarized gravitational wave composed of two basis vector modes
with a waveform described by a simple sine-Gaussian wavepacket, with some
characteristic frequency $\Omega $ and relaxation time $\tau $. Then the
strain measured by a given detector $I$ will be \cite{Isi17}:%
\begin{equation}
h_{I}(t)=\text{Re}\left[ A\left( F_{1}^{I}+i\epsilon F_{2}^{I}\right)
e^{i\Omega (t-t_{I})}\right] e^{-\left( t-t_{I}\right) ^{2}/\tau ^{2}},
\label{w1}
\end{equation}%
where $F_{1}^{I}$ and $F_{2}^{I}$ are the responses of the detector $I$ to
the two basis polarizations, $A\equiv |A|e^{i\phi _{0}}$ is a complex-valued
amplitude, $\epsilon $ is an ellipticity parameter controlling the relative
amounts of each polarization and $t_{I}$ is the arrival time of the wave at
the location of the interferometer $I$ ($I=H$, $L$, $V$). Parameters $%
F_{1}^{I}$ and $F_{2}^{I}$ are determined by the propagation direction of
the gravitational wave and orientation of the interferometer arms. If
gravitational wave propagates along the unit vector $\hat{k}$ then in vector
gravity the unit basis polarization vectors $\mathbf{\hat{e}}_{1}$ and $%
\mathbf{\hat{e}}_{2}$ are perpendicular to $\hat{k}$ \cite{Svid17}. They can
be chosen in any convenient way, e.g., such that $\mathbf{\hat{e}}_{1}\times 
\mathbf{\hat{e}}_{2}=\hat{k}$. Assuming that arms of the detector $I$ point
along unit vectors $\hat{a}_{I}$ and $\hat{b}_{I}$ the detector response
functions read 
\begin{equation}
F_{1}^{I}=(\hat{a}_{I}\cdot \hat{k})(\hat{a}_{I}\cdot \mathbf{\hat{e}}_{1})-(%
\hat{b}_{I}\cdot \hat{k})(\hat{b}_{I}\cdot \mathbf{\hat{e}}_{1}),  \label{f1}
\end{equation}%
\begin{equation}
F_{2}^{I}=(\hat{a}_{I}\cdot \hat{k})(\hat{a}_{I}\cdot \mathbf{\hat{e}}_{2})-(%
\hat{b}_{I}\cdot \hat{k})(\hat{b}_{I}\cdot \mathbf{\hat{e}}_{2}).  \label{f2}
\end{equation}

\begin{figure}[t]
\begin{center}
\epsfig{figure=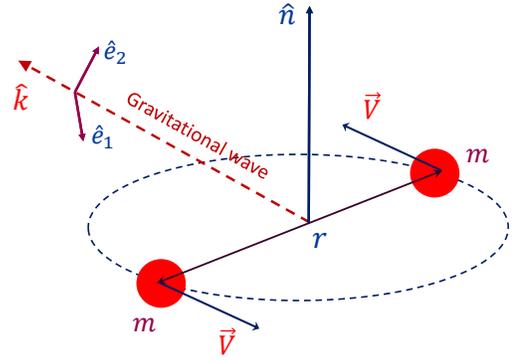, angle=270, width=8cm}
\end{center}
\par
\vspace{-0.5cm}
\caption{Two neutron stars of mass $m$ move along a circular orbit of
diameter $r$ emitting gravitational waves. Unit vector $\hat{n}$ is
perpendicular to the orbital plane. Propagation direction of the
gravitational wave is given by the unit vector $\hat{k}$. $\mathbf{\hat{e}}%
_{1}$ and $\mathbf{\hat{e}}_{2}$ are unit polarization vectors perpendicular
to $\hat{k}$.}
\label{Orientation}
\end{figure}

Isi and Weinstein assume that ellipticity parameter $\epsilon $ in Eq. (\ref%
{w1}) is a real number and write Eq. (\ref{w1}) in the form%
\begin{equation}
h_{I}(t)=A_{I}\cos \left[ \Omega (t-t_{I})+\Phi _{I}\right] e^{-\left(
t-t_{I}\right) ^{2}/\tau ^{2}},  \label{f3}
\end{equation}%
where 
\begin{equation}
A_{I}=|A|\left\vert F_{1}^{I}+i\epsilon F_{2}^{I}\right\vert ,  \label{f4}
\end{equation}%
\begin{equation}
\Phi _{I}=\phi _{0}+\arctan \left( \epsilon F_{2}^{I}/F_{1}^{I}\right)
\label{f5}
\end{equation}%
are the main observables at each detector. They can be obtained, along with $%
t_{I}$, by fitting output of each detector with the waveform (\ref{f3}). The
three timing measurements $t_{I}$ are sufficient to find the propagation
direction of the gravitational wave $\hat{k}$. This information, according
to Eqs. (\ref{f1}) and (\ref{f2}), will then fix the values of $F_{1}^{I}$
and $F_{2}^{I}$. However, since the two LIGO instruments (in Hanford and
Livingston) are almost co-aligned, they have the same response functions: $%
F_{1}^{H}=F_{1}^{L}$, $F_{2}^{H}=F_{2}^{L}$ and Eqs. (\ref{f4}) and (\ref{f5}%
) yield $A_{H}=A_{L}$, $\Phi _{H}=\Phi _{L}$. Thus, we are left with only
four independent equations 
\begin{equation}
A_{L}=|A|\left\vert F_{1}^{L}+i\epsilon F_{2}^{L}\right\vert ,  \label{f6}
\end{equation}%
\begin{equation}
A_{V}=|A|\left\vert F_{1}^{V}+i\epsilon F_{2}^{V}\right\vert ,  \label{f7}
\end{equation}%
\begin{equation}
\Phi _{L}=\phi _{0}+\arctan \left( \epsilon F_{2}^{L}/F_{1}^{L}\right) ,
\label{f8}
\end{equation}%
\begin{equation}
\Phi _{V}=\phi _{0}+\arctan \left( \epsilon F_{2}^{V}/F_{1}^{V}\right) ,
\label{f9}
\end{equation}%
for the three unknown fitting parameters $|A|$, $\phi _{0}$ and $\epsilon $.
Hence, the system of equations (\ref{f6})-(\ref{f9}) is overdetermined
(there are more equations than unknowns). In general case such a system does
not have solutions, which means that it would be impossible to fit the
measured waveform unless gravitational wave really has vector polarization.
Pure vector polarization of gravitational waves will be compatible with
observations only if, subject to the experimental uncertainties, Eqs. (\ref%
{f6})-(\ref{f9}) have solutions for $|A|$, $\phi _{0}$ and $\epsilon $.
Based on this analysis, Isi and Weinstein concluded that a single detection
event by the LIGO-Virgo network can distinguish between pure tensor and pure
vector polarizations of gravitational waves.

\begin{figure}[t]
\begin{center}
\epsfig{figure=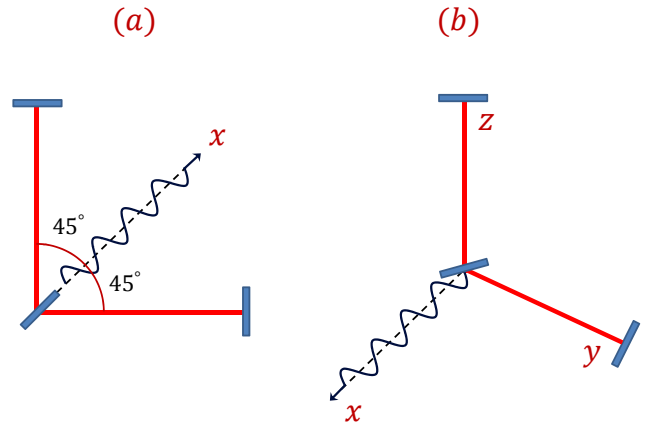, angle=270, width=9cm}
\end{center}
\par
\vspace{-1cm}
\caption{(a) Gravitational wave propagates in the interferometer plane at $%
45^{\circ } $ angle relative to the interferometer arms. Such wave produces
no signal in general relativity. (b) Gravitational wave propagates in the
direction perpendicular to the interferometer plane. Such wave yields no
signal in vector gravity.}
\label{Arms}
\end{figure}

Such analysis, however, is erroneous because parameter $\epsilon $ in Eq. (%
\ref{w1}) is mistakenly taken as a real number. Instead, it must be taken as
a complex number (see Appendix A)%
\begin{equation}
\epsilon =\epsilon _{1}+i\epsilon _{2},
\end{equation}%
where 
\begin{equation}
\epsilon _{1}=\frac{(\hat{k}\cdot \hat{n})}{1-(\mathbf{\hat{e}}_{1}\cdot 
\hat{n})^{2}},\quad \epsilon _{2}=\frac{(\mathbf{\hat{e}}_{1}\cdot \hat{n})(%
\mathbf{\hat{e}}_{2}\cdot \hat{n})}{1-(\mathbf{\hat{e}}_{1}\cdot \hat{n})^{2}%
}.
\end{equation}%
Here $\hat{n}$ is a unit vector perpendicular to the orbital plane of the
binary system which describes orientation of this plane (see Fig. \ref%
{Orientation}). Direction of $\hat{n}$ is unknown. It is described by two
angles and, thus, $\epsilon _{1}$ and $\epsilon _{2}$ are independent
fitting parameters. As a result, Eqs. (\ref{f6})-(\ref{f9}) must be replaced
with four equations 
\begin{equation}
A_{L}=|A|\left\vert F_{1}^{L}-\epsilon _{2}F_{2}^{L}+i\epsilon
_{1}F_{2}^{L}\right\vert ,
\end{equation}%
\begin{equation}
A_{V}=|A|\left\vert F_{1}^{V}-\epsilon _{2}F_{2}^{V}+i\epsilon
_{1}F_{2}^{V}\right\vert ,
\end{equation}%
\begin{equation}
\Phi _{L}=\phi _{0}+\arctan \left( \frac{\epsilon _{1}F_{2}^{L}}{%
F_{1}^{L}-\epsilon _{2}F_{2}^{L}}\right) ,  \label{s6}
\end{equation}%
\begin{equation}
\Phi _{V}=\phi _{0}+\arctan \left( \frac{\epsilon _{1}F_{2}^{V}}{%
F_{1}^{V}-\epsilon _{2}F_{2}^{V}}\right) ,  \label{s7}
\end{equation}%
for the four unknown fitting parameters $|A|$, $\phi _{0}$, $\epsilon _{1}$
and $\epsilon _{2}$. Such system of equations is no longer overdetermined
and in most instances has solutions. That is no matter what are the measured
values of $A_{I}$ and $\Phi _{I}$ the measured waveform can be fitted by the
gravitational waves of vector polarization. Therefore, in most instances, a
single detection event by the LIGO-Virgo network will be compatible with the
pure vector and pure tensor polarizations.

Differentiation between the two polarizations can be made based on a single
detection only in the unlikely event when gravitational wave propagates in
the direction for which%
\begin{equation}
F_{1}^{I}=F_{2}^{I}=0
\end{equation}%
for one of the interferometers. Such directions of zero response are
different for tensor and vector polarizations \cite{Svid17}. Namely, tensor
gravitational wave produces no signal when it propagates parallel to the
interferometer plane at $45^{\circ }$ angle relative to one of the
perpendicular arms (see Fig. \ref{Arms}\textit{a}). On the other hand,
vector gravitational wave yields no signal if it propagates in the direction
perpendicular to the interferometer plane (see Fig. \ref{Arms}\textit{b}),
or along one of the interferometer arms. One should mention that propagation
direction perpendicular to the interferometer plane is also the direction of
zero response for the longitudinal (scalar) gravitational waves which also
exist in the vector theory of gravity \cite{Svid17}.

Detection of a gravitational wave in the direction of zero response can rule
out pure tensor or pure vector polarizations based on a single event.
However, wave propagation direction can be obtained accurately only from
observing the source electromagnetic counterpart which further reduces
chances of making a decisive conclusion based on a single detection. So far
the source of gravitational waves was found only for the signal GW170817 
\cite{Abbo17a}. For this event, the gravitational wave propagated about $%
20^{\circ }$ off the nearest direction of zero response for the tensor
polarization and about $30^{\circ }$ for the vector polarization. Thus, no
conclusion in favor of one of the two polarizations can be made.

The LIGO-Virgo network can distinguish between the two polarizations by
collecting statistics which should exhibit minima in the distribution
function in the vicinities of the directions of the zero response. Tensor
and vector polarizations predict that such minima occur at different
positions. See Section 16 and Figure 10 in \cite{Svid17} for details.

This work was supported by the Office of Naval Research (Award Nos.
N00014-16-1-3054 and N00014-16-1-2578), and the Robert A. Welch Foundation
(Award A-1261).

\appendix

\section{Radiation of gravitational waves in vector gravity}

For a weak transverse gravitational wave propagating along the $x-$axis in
vector gravity the equivalent metric evolves as \cite{Svid17}%
\begin{equation}
g_{ik}=\eta _{ik}+\left( 
\begin{array}{cccc}
0 & 0 & h_{0y}(t,x) & h_{0z}(t,x) \\ 
0 & 0 & 0 & 0 \\ 
h_{0y}(t,x) & 0 & 0 & 0 \\ 
h_{0z}(t,x) & 0 & 0 & 0%
\end{array}%
\right) ,  \label{met1}
\end{equation}%
where $\eta _{ik}$ is Minkowski metric. The three-dimensional polarization
vector of this wave is 
\begin{equation}
\mathbf{h}=(0,h_{0y},h_{0z}).
\end{equation}

Here we consider a general case when gravitational wave propagates along the
unit vector $\hat{k}$. Then polarization vector $\mathbf{h}$ is
perpendicular to $\hat{k}$ so that $\hat{k}\cdot \mathbf{h}=0$.

Signal of the LIGO-like interferometer with perpendicular arms of length $L$
along the direction of unit vectors $\hat{a}$ and $\hat{b}$ is proportional
to the relative phase shift $\Delta \varphi $ of electromagnetic waves
traveling a roundtrip distance $2L$ along the two arms \cite{Svid17} 
\begin{equation}
\Delta \varphi =\frac{2\omega L}{c}\left[ (\hat{a}\cdot \hat{k})(\hat{a}%
\cdot \mathbf{h})-(\hat{b}\cdot \hat{k})(\hat{b}\cdot \mathbf{h})\right] ,
\label{wa4}
\end{equation}%
where $\omega $ is the frequency of electromagnetic wave. Eq. (\ref{wa4})
shows that for any $\mathbf{h}$ the interferometer has zero response if
gravitational wave propagates perpendicular to the interferometer plane [$%
\hat{a}\cdot \hat{k}=\hat{b}\cdot \hat{k}=0$] or along one of the
interferometer arms. This result is independent of polarization of the
transverse gravitational wave and can rule out vector theory of gravity if
gravitational wave is detected propagating in the predicted direction of the
zero response.

Depending on the inclination angle between $\hat{k}$ and the orbital plane
of binary stars, gravitational wave in vector gravity can be linearly or
elliptically polarized in the same way as electromagnetic wave generated by
an oscillating quadrupole. Recall that in vector gravity $\mathbf{h}$ is
analogous to the vector potential $\mathbf{A}$ in electrodynamics \cite%
{Svid17}.

To be specific, we consider generation of gravitational waves by two compact
stars with equal masses $m$ moving along circular orbits of diameter $r$
with angular velocity $\Omega $ (see Fig. \ref{Orientation}). $\theta
(t)=\Omega t+\theta _{0}$ is the star azimuthal angle in the orbital plane.
Using formulas of Ref. \cite{Svid17} we obtain that far from the binary
system 
\begin{equation}
\mathbf{h}=-\frac{2Gm\Omega ^{2}r^{2}}{c^{4}R}\text{Re}\left[ \mathbf{P}%
e^{2i\theta (t)}\right] ,  \label{s1}
\end{equation}%
where $R$ is the distance to the system, $\mathbf{P}$ is the complex
polarization vector 
\begin{equation}
\mathbf{P}=(\hat{k}\cdot \hat{n})\left[ \hat{k}\times \lbrack \hat{k}\times 
\hat{n}]\right] -i[\hat{k}\times \hat{n}],  \label{s2}
\end{equation}%
and $\hat{n}$ is a unit vector perpendicular to the star orbital plane (see
Fig. \ref{Orientation}).

We denote real unit polarization vectors as $\mathbf{\hat{e}}_{1}$ and $%
\mathbf{\hat{e}}_{2}$. They are perpendicular to $\hat{k}$ and can be chosen
such that $\mathbf{\hat{e}}_{1}\times \mathbf{\hat{e}}_{2}=\hat{k}$. Using
Eq. (\ref{s1}) one can write $\mathbf{h}$ as%
\begin{equation}
\mathbf{h}=-\frac{2Gm\Omega ^{2}r^{2}}{c^{4}R}\text{Re}\left[ \left( (%
\mathbf{P\cdot \hat{e}}_{1})\mathbf{\hat{e}}_{1}+(\mathbf{P\cdot \hat{e}}%
_{2})\mathbf{\hat{e}}_{2}\right) e^{2i\theta (t)}\right] ,  \label{s3}
\end{equation}%
Then Eq. (\ref{wa4}) yields the following expression for the interferometer
response%
\begin{equation}
\Delta \varphi =-\frac{4G\omega Lm\Omega ^{2}r^{2}}{c^{5}R}\text{Re}\left[
\left( (\mathbf{P\cdot \hat{e}}_{1})F_{1}+(\mathbf{P\cdot \hat{e}}%
_{2})F_{2}\right) e^{2i\theta (t)}\right] ,  \label{s4}
\end{equation}%
where $F_{1}$ and $F_{2}$ are the detector response functions to the
polarizations $\mathbf{\hat{e}}_{1}$ and $\mathbf{\hat{e}}_{2}$ 
\begin{equation}
F_{1,2}=(\hat{a}\cdot \hat{k})(\hat{a}\cdot \mathbf{\hat{e}}_{1,2})-(\hat{b}%
\cdot \hat{k})(\hat{b}\cdot \mathbf{\hat{e}}_{1,2})
\end{equation}%
and 
\begin{equation}
(\mathbf{P\cdot \hat{e}}_{1})=-(\hat{k}\cdot \hat{n})(\mathbf{\hat{e}}%
_{1}\cdot \hat{n})+i(\mathbf{\hat{e}}_{2}\cdot \hat{n}),
\end{equation}%
\begin{equation}
(\mathbf{P\cdot \hat{e}}_{2})=-(\hat{k}\cdot \hat{n})(\mathbf{\hat{e}}%
_{2}\cdot \hat{n})-i(\mathbf{\hat{e}}_{1}\cdot \hat{n}).
\end{equation}

Comparing Eq. (\ref{s4}) with Eq. (\ref{w1}) we find that parameter $%
\epsilon $ is given by 
\begin{equation}
i\epsilon =\frac{(\mathbf{P\cdot \hat{e}}_{2})}{(\mathbf{P\cdot \hat{e}}_{1})%
}
\end{equation}%
or 
\begin{equation}
\epsilon =\frac{(\hat{k}\cdot \hat{n})+i(\mathbf{\hat{e}}_{1}\cdot \hat{n})(%
\mathbf{\hat{e}}_{2}\cdot \hat{n})}{1-(\mathbf{\hat{e}}_{1}\cdot \hat{n})^{2}%
}.
\end{equation}

When $\hat{k}$ is parallel to the orbital plane ($\hat{k}\cdot \hat{n}=0$)
the gravitational wave is linearly polarized and polarization vector $%
\mathbf{h}$ is parallel to the orbital plane. In this case $\epsilon $ is
pure imaginary and, according to Eqs. (\ref{s6}) and (\ref{s7}), $\Phi
_{L}=\Phi _{V}$. That is phase shift of the detected signal between
different interferometers is determined only by the time delay. Since
orientation of the orbital plane is unknown, the direction of $\mathbf{h}$
is a free parameter. Using vector identities one can obtain from Eq. (\ref%
{wa4}) that for the plane gravitational wave the interferometer signal is
equal to zero if $\mathbf{h}$ is parallel to the vector 
\begin{equation}
\mathbf{h}\parallel \hat{k}\times \left[ (\hat{a}\cdot \hat{k})\hat{a}-(\hat{%
b}\cdot \hat{k})\hat{b}\right] .  \label{a1}
\end{equation}%
So, if for fixed $\hat{k}$ the wave polarization $\mathbf{h}$ satisfies Eq. (%
\ref{a1}) for the orientation of the Virgo (LIGO) interferometer arms then
Virgo (LIGO) signal will be zero. At the same time, orientation of the LIGO
(Virgo) arms will not satisfy Eq. (\ref{a1}) and the signal can be nonzero.
Thus, the ratio of the LIGO/Virgo signal amplitudes depends on the unknown
polarization direction $\mathbf{h}$. For fixed $\hat{k}$, depending on the
direction of $\mathbf{h}$, the LIGO/Virgo signal ratio can be any number
between$\ 0$ and $\infty $.

If inclination angle is nonzero, then gravitational wave will be
elliptically polarized. In this case $\epsilon $ has both real and imaginary
parts and $\Phi _{L}\neq \Phi _{V}$. Degree of ellipticity determines the
extra phase shift between LIGO and Virgo signals which is not caused by the
difference in the arrival times of the wave at the interferometer locations.
The inclination angle is also a free parameter. In total, the orientation of
the orbital plane of the binary system is described by two free parameters
(angles of $\hat{n}$) which determine the ratio of the LIGO/Virgo signal
amplitudes and the extra phase shift between the measured waveforms.


\begin{thebibliography}{9}
\bibitem{Abbo17} B.\thinspace P. Abbott et al. (LIGO Scientific
Collaboration and Virgo Collaboration), \textit{GW170814: A Three-Detector
Observation of Gravitational Waves from a Binary Black Hole Coalescence},
Phys. Rev. Lett. \textbf{119}, 141101 (2017).

\bibitem{Abbo17a} B.\thinspace P. Abbott et al. (LIGO Scientific
Collaboration and Virgo Collaboration), \textit{GW170817: Observation of
Gravitational Waves from a Binary Neutron Star Inspiral}, Phys. Rev. Lett. 
\textbf{119}, 161101 (2017).

\bibitem{Eins15} A. Einstein, \textit{Die Feldgleichungen der Gravitation, }%
Sitzungsber. Preuss. Akad. Wiss.\textit{\ }\textbf{1915}, 844 (1915).

\bibitem{Svid17} A.A. Svidzinsky, \textit{Vector theory of gravity: Universe
without black holes and solution of dark energy problem}, Phys. Scr. \textbf{%
92,} 125001 (2017).

\bibitem{Isi17} M. Isi and A.J. Weinstein, \textit{Probing gravitational
wave polarizations with signals from compact binary coalescences},
arXiv:1710.03794v1 [gr-qc].
\end{thebibliography}
\end{document}